\documentclass[]{pasj01}

\begin{document} 
\Received{}
\Accepted{}

\title{Study of Nature of Corona in Narrow Line Seyfert 1 Galaxies}

\author{Hiroshi H. \textsc{Idogaki}\altaffilmark{1}$^{*}$}
\author{Shunichi \textsc{Ohmura}\altaffilmark{1}}
\author{Hiroaki \textsc{Takahashi}\altaffilmark{2}}
\author{Shogo B. \textsc{Kobayashi}\altaffilmark{3}}
\author{Yoshihiro \textsc{Ueda}\altaffilmark{1}}
\author{Yuichi \textsc{Terashima}\altaffilmark{4}}
\author{Kiyoshi \textsc{Hayashida}\altaffilmark{2}}
\author{Takaaki \textsc{Tanaka}\altaffilmark{1}}
\author{Hiroyuki \textsc{Uchida}\altaffilmark{1}}
\author{Takeshi Go \textsc{Tsuru}\altaffilmark{1}}

\altaffiltext{1}{Department of Physics, Kyoto University, Kitashirakawa Oiwake-cho, Sakyo-ku, Kyoto 606-8502, Japan}
\altaffiltext{2}{Department of Earth and Space Science, Osaka University, 1-1 Machikaneyama-cho, Toyonaka, Osaka 560-0043, Japan}
\altaffiltext{3}{Department of Physics, Tokyo University of Science, 1-3 Kagurazaka, Shinjuku-ku, Tokyo 162-8601, Japan}
\altaffiltext{4}{Department of Physics, Ehime University, 2-5 Bunkyo-cho, Matsuyama, Ehime, 790-8577, Japan}
\email{idogaki.hiroshi.47m@kyoto-u.jp}


\KeyWords{galaxies: active --- galaxies: Seyfert --- X-rays: galaxies} 

\maketitle

\begin{abstract}
We study X-ray spectra of four Narrow Line Seyfert 1 galaxies (NLS1s), Mrk~110, SWIFT~J2127.4+5654, IGR~J16185$-$5928, and WKK~4438, using Suzaku and NuSTAR data. 
The spectra of the four sources are reproduced with a model consisting of a cutoff power law component, a reflection component, and a soft excess component. 
The photon indices of all the sources are found to be $\sim 2$. 
The cutoff energies are constrained to be $\sim 40~\mathrm{keV}$ for Mrk~110, SWIFT~J2127.4+5654, and WKK~4438, whereas a lower limit of $155~\mathrm{keV}$ is obtained for IGR~J16185$-$5928. 
We find that the NLS1s in our sample have systematically softer spectra and lower cutoff energies than Broad Line Seyfert 1 galaxies reported in the literature. 
We also perform spectral fittings with a model in which the cutoff power law is replaced with the thermal Comptonization model in order to directly obtain the electron temperature of the corona ($kT_{\mathrm{e}}$). 
The fits give $kT_{\mathrm{e}} \sim 10$--$20~\mathrm{keV}$ for Mrk~110 and SWIFT~J2127.4+5654, and lower limits of $kT_{\mathrm{e}} > 18~\mathrm{keV}$ for IGR~J16185$-$5928 and $kT_{\mathrm{e}} > 5~\mathrm{keV}$ for WKK~4438. 
We divide the NuSTAR data of SWIFT~J2127.4+5654 into three periods according to the flux. 
The spectral fits in the three periods hint that $kT_{\mathrm{e}}$ is the lower in the higher flux period. 
The results of the four sources suggest a possible anti-correlation between $kT_{\mathrm{e}}$ and the ratio of the luminosity of Compton up-scattered photons to the Eddington luminosity. 
The anti-correlation can be explained by a simple model in which electrons in the corona are cooled and heated through inverse Compton scattering and Coulomb collisions with protons, respectively. 

\end{abstract}

\section{Introduction}
\begin{table*}
  \tbl{Basic parameters of targets.}{%
  \begin{tabular}{ccccc}
      \hline
      Object & Mrk~110 & SWIFT~J2127.4+5654 & IGR~J16185$-$5928 & WKK~4438 \\
      \hline
      Coordinates (J2000) & (09$^{\rm{h}}$25$^{\rm{m}}$12$^{\rm{s}}$9, +52$^{\circ}$17$^{\prime}$11$^{\prime\prime}$) & (21$^{\rm{h}}$27$^{\rm{m}}$44$^{\rm{s}}$9, +56$^{\circ}$56$^{\prime}$40$^{\prime\prime}$) & (16$^{\rm{h}}$18$^{\rm{m}}$36$^{\rm{s}}$4, -59$^{\circ}$27$^{\prime}$17$^{\prime\prime}$)　& (14$^{\rm{h}}$55$^{\rm{m}}$16$^{\rm{s}}$7, -51$^{\circ}$34$^{\prime}$16$^{\prime\prime}$) \\
      redshift $z$ & 0.0353 & 0.0147 & 0.0356 & 0.0160 \\
      $N_{\rm{H}}$~($\times 10^{22}~\mathrm{cm}^{-2}$)\footnotemark[$*$] & 0.0133 & 0.770 & 0.216 & 0.291 \\
      $M_{\rm{BH}}$~($\times 10^7 M_{\mathrm{\odot}}$)\footnotemark[$\dagger$] & 1.8\footnotemark[$\ddagger$] & 1.5\footnotemark[$\S$] & 2.8\footnotemark[$\|$] & 0.2\footnotemark[$\|$] \\
      \hline
    \end{tabular}}\label{tab:basic}
\begin{tabnote}
	\footnotemark[$*$] the column density of Galactic interstellar medium\\
	\footnotemark[$\dagger$] the mass of SMBH\\
	\footnotemark[$\ddagger$] \citet{kollatschny2004}~
	\footnotemark[$\S$] \citet{malizia2008}~
	\footnotemark[$\|$] \citet{masetti2006} \\
\end{tabnote}
\end{table*}

\begin{table*}
  \tbl{Obserbvation logs.}{%
  \begin{tabular}{lcccc}
      \hline
      Object & Date & ID & Exposure (ks) & Observatory \\ 
      \hline
      Mrk~110 & 2007/11/02 & 702124010 & 80 & Suzaku \\
      SWIFT~J2127.4+5654 & 2007/12/09 & 702122010 & 82 &  Suzaku \\
       & 2012/11/04 & 60001110002 & 49 & NuSTAR \\
       & 2012/11/05 & 60001110003 & 29 & NuSTAR \\
       & 2012/11/06 & 60001110005 & 75 & NuSTAR \\
       & 2012/11/08 & 60001110007 & 42 & NuSTAR  \\
      IGR~J16185$-$5928 & 2008/02/09 & 702123010 & 68 & Suzaku \\
      WKK~4438 & 2012/01/22 & 706011010 & 70 & Suzaku \\
      \hline
    \end{tabular}}\label{obs}
\end{table*}

Seyfert 1 galaxies (Sy1s) host Active Galactic Nuclei (AGNs), and are divided into two subclasses, Narrow Line Seyfert 1 galaxies (NLS1s) and Broad Line Seyfert 1 galaxies (BLS1s), based on the width of the $\mathrm{H}\beta$ emission line \citep{osterbrock1985}.
According to the studies by e.g., \citet{jin2012} and \citet{ai2013}, NLS1s and BLS1s have similar bolometric luminosities $L_{\mathrm{bol}}$ ($\sim 10^{44-47}~\mathrm{erg~s^{-1}}$) 
although the SMBH mass of NLS1s is typically smaller ($M \sim 10^{6-7}M_{\mathrm{\odot}}$).
Thus, the Eddington ratios of NLS1s are generally higher than those of BLS1s, implying that NLS1s are in the early phase of the Super Massive Black Hole (SMBH) growth.

X-ray spectra of Sy1s are described roughly by a power law with a high energy cutoff at $\sim 10^{1-2}~\mathrm{keV}$. 
The emission mechanism is supposed to be inverse Compton scattering, in which low energy photons from the accretion disk are up-scattered by thermal electrons in the accretion corona \citep{haardt1994}. 
Some of the up-scattered photons are reflected (Compton down-scattered and photoelectrically absorbed) by surrounding matter \citep{george1991} such as the dusty torus and/or the disk. 
These reflected photons produce the common features such as the Compton hump  as well as the Fe K$\alpha$ emission line and the Fe K$\alpha$ edge.

\citet{malizia2003} studied hard X-ray spectra of nine BLS1s, analyzing data obtained with the BeppoSAX MECS and PDS. 
They combined spectra of the nine BLS1s, and obtained a cutoff energy of $E_{\mathrm{c}} = 216^{+75}_{-41}~\mathrm{keV}$.
\citet{malizia2008} carried out a similar analysis for five NLS1s with the Swift XRT and the INTEGRAL IBIS, and obtained $E_{\mathrm{c}} = 38^{+17}_{-10}~\mathrm{keV}$. 
Since $E_{\mathrm{c}} \approx 2$--$3\, kT_{\mathrm{e}}$ \citep{petrucci2001}, the results by Malizia et al.\ (\yearcite{malizia2003}, \yearcite{malizia2008}) imply that NLS1s have lower electron temperatures than BLS1s. \citet{malizia2008} claimed that the higher accretion rate of NLS1s enhances Compton up-scattering, and that the corona of NLS1s dose not have a temperature as high as that of BLS1s. 

For individual sources, \citet{malizia2014} performed 0.3--100~keV spectral analysis of 41 BLS1s included in the INTEGRAL complete sample of AGNs. 
They had 26 measurements of the cutoff distributed between 50 and 200 keV with the mean value of the cutoff energies of $E_{\mathrm{c}} = 128~\mathrm{keV}$, which corresponds to $kT_{\mathrm{e}} \sim 50~\mathrm{keV}$. 
Although these measurements were all assuming a phenomenological cutoff power law model to determine the spectral cutoff, a recent study by \citet{tortosa2018} directly determined the electron temperatures of two BLS1s, MCG~+8$-$11$-$11 and NGC~6814, with NuSTAR by utilizing a Comptonization model as $60^{+110}_{-30}~\mathrm{keV}$ and $45^{+100}_{-17}~\mathrm{keV}$, respectively. 
As for NLS1s, several studies have been carried out to characterize the properties of individual sources in the hard X-ray band (\cite{malizia2008}; \cite{kara2017}; \cite{marinucci2014}; \cite{miniutti2009}; \cite{patrick2011}; \cite{panessa2011}). 
In addition to well-studied SWIFT~J2127.4+5654, cutoff energies of three NLS1s have been successfully determined so far by those authors. 

In this paper, we study spectra of four NLS1s obtained with Suzaku \citep{mitsuda2007} and NuSTAR \citep{harrison2013} in order to measure their cutoff energies and characterize the properties of the corona. 
Throughout this paper, the Hubble constant $H_{0} = 70~\mathrm{km~s^{-1}~Mpc^{-1}}$ is adopted.
Errors are quoted at 90\% confidence levels in the text and tables, and error bars in the figures indicate $1\sigma$ confidence intervals.

\section{Observations and Data Reduction}
As the targets of the present study, we selected Mrk~110, SWIFT~J2127.4+5654, IGR~J16185$-$5928, and WKK~4438, which have relatively high flux ($\gtrsim 10^{-11}~\mathrm{erg~s^{-1}~cm^{-2}}$ in 14--195~keV band) in the Swift BAT 58 months catalog \citep{baumgartner2010} to study the properties in the hard X-ray band and obtain cutoff energies. 
Their basic parameters and the observation logs are summarized in tables~\ref{tab:basic} and \ref{obs}, respectively.

\subsection{Suzaku Observations}

The Suzaku observations were carried out with the X-ray Imaging Spectrometer (XIS; \cite{koyama2007}) and the Hard X-ray Detector (HXD; \cite{takahashi2007}; \cite{kokubun2007}).
The XIS consists of four CCD cameras: three (XIS0, XIS2, XIS3) with front-illuminated (FI) CCDs and one (XIS1) with a back-illuminated (BI) CCD. 
We did not use XIS2 because it had been out of function since 2006.
The XIS0 data of  SWIFT~J2127.4+5654 are not available because of a trouble in on-board data processing.
The normal full-window clocking mode with the spaced-row charge injection (\cite{ozawa2009}; \cite{uchiyama2009}) was employed. 

We reprocessed the XIS and HXD datasets by using the  Calibration Database (CALDB) released in 2016 April and reduced them by using the HEASOFT package version 6.18.
We employed the same data-selection criteria as those used for the archived cleaned event lists.
We analyzed the data in the time intervals in which both the XIS and HXD data were simultaneously available.

Each XIS source spectrum was extracted from circular regions with a 3$^{\prime}$ radius centered at the source.
The background was taken from off-source circular regions with a 3$^{\prime}$ radius for Mrk~110, SWIFT~J2127.4+5654, and IGR~J16185$-$5928, which were observed at the HXD nominal position. 
WKK~4438 was observed at the XIS nominal position, which makes it difficult to select a background region in the above mentioned manner. 
Thus, the background for WKK~4438 was taken from an annular region with inner and outer radii of 3$^{\prime}$ and 6$^{\prime}$, respectively. 
The redistribution matrices and the ancillary response files were made for each XIS spectra using {\tt xisrmfgen} and {\tt xissimarfgen}, respectively. 

We followed the standard data reduction procedure for the HXD-PIN data.
For the HXD-PIN Non X-ray Background (NXB), we used simulated event files supplied by the HXD team based on the LCFITDT model \citep{fukazawa2009},
which is estimated to have $1\sigma$ systematic errors less than $3\%$ in the 15--40~keV band. 
The Cosmic X-ray Background (CXB) component was simulated by assuming the empirical model by \citet{boldt1987}.
We simulated the CXB component with this spectral model and the HXD response file for a uniform diffuse emission, and added to the NXB component to make the total background spectrum.
In the spectral analysis, we used the HXD-PIN response files, ae\_hxd\_pinhxnome4\_20080129.rsp and ae\_hxd\_pinxinome11\_20110601.rsp for the HXD nominal position and XIS nominal position, respectively.

\subsection{NuSTAR Observation}
The NuSTAR observation of SWIFT~J2127.4+5654 was carried out with Focal Plane Module A (FPMA) and B (FPMB), which consist of CdZnTe pixel detectors. We reprocessed the data by using the CALDB released in 2017 May and reduced them by using the HEASOFT package version 6.21. We then extracted high-level scientific products from cleaned Level 2 ``01'' event files by using {\tt nuproducts}.
The source spectra were extracted from a circular region with a 1.9$^{\prime}$ radius centered at the source, and the background was extracted from an off-source circular region with a 1.9$^{\prime}$ radius.

\section{Spectral Analysis and Results}
\subsection{Suzaku Analysis}\label{suzakuana}
\begin{figure*}
 \begin{center}
  \includegraphics[width=15cm]{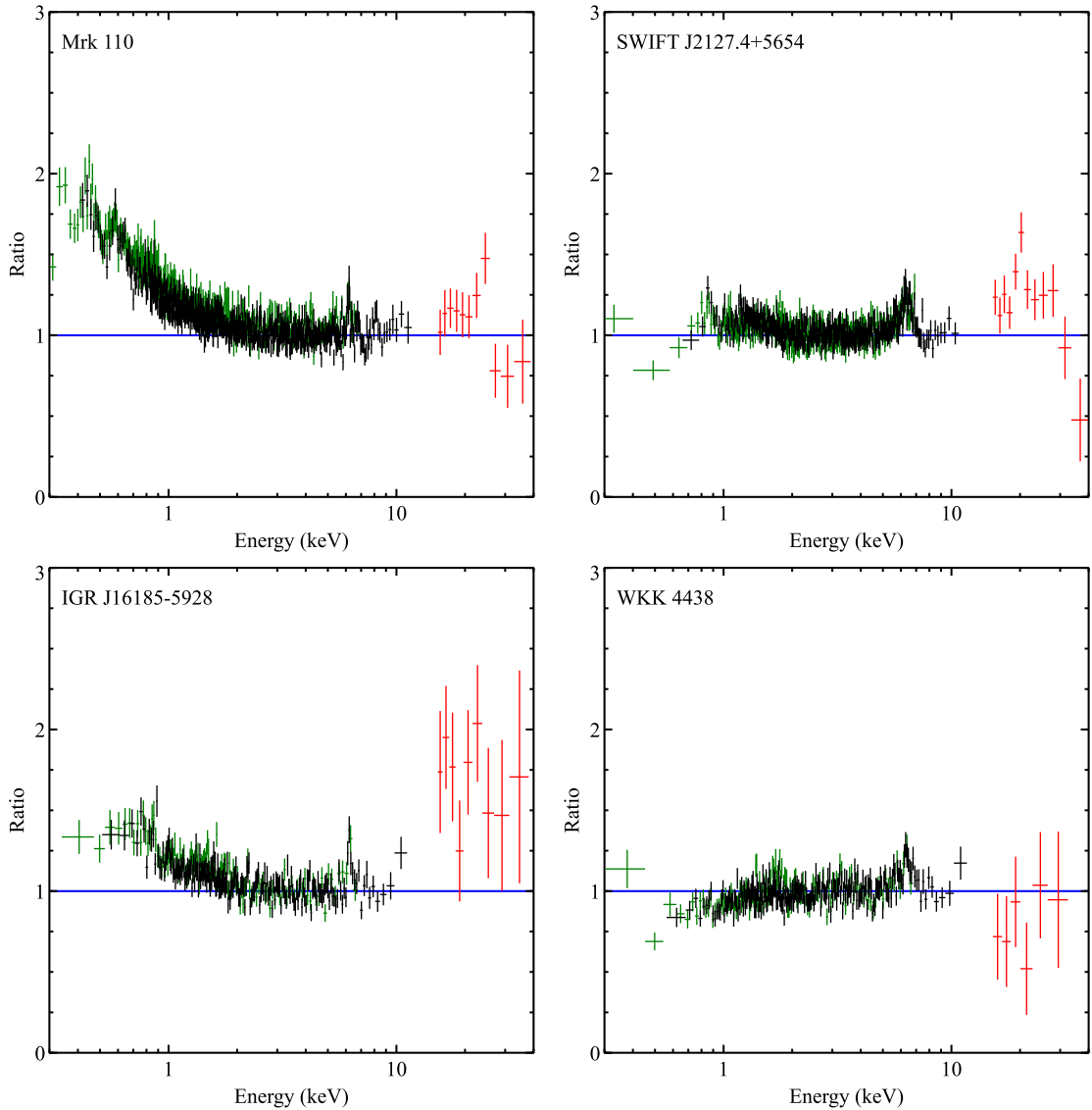} 
 \end{center}
\caption{Ratios of the spectra of Mrk~110, SWIFT~J2127.4+5654, IGR~J16185$-$5928, and WKK~4438 to power law models. The black, green, and red crosses correspond to FI-XIS, BI-XIS, and HXD-PIN, respectively.}\label{ratio}
\end{figure*}

We analyzed time averaged spectra of the four NLS1s by using XSPEC \citep{arnaud1996}. 
The cross-normalization factor of the HXD-PIN with respect to FI-XIS was set to 1.16 for the XIS nominal position observation and 1.18 for the HXD nominal position observations by referring to Suzaku Memo\footnote{$\langle$ftp://legacy.gsfc.nasa.gov/suzaku/doc/xrt/suzakumemo-2008-06.pdf$\rangle$.}. 
In order to grasp overall spectral shape, we first used a single power law model absorbed by interstellar medium ({\tt phabs} in XSPEC). 
In order to avoid the influence of the structures commonly seen in the spectra of NLS1s such as the soft excess, the Compton hump, and the Fe K$\alpha$ emission line, the model was applied to the 2.5--12~keV energy band excluding 4.5--7.5~keV.  
The absorption column densities were fixed to the Galactic values taken from the results of the Leiden/Argentine/Bonn Survey \citep{kalberla2005} in the following analyses. 
Figure~\ref{ratio} shows the spectral ratios between the data in the 0.3--40~keV energy band and the model. 
The fits gave photon indices $\Gamma$ in the range of 1.7--2.0. We found cutoff features around 20~keV in the spectra of Mrk~110, SWIFT~J2127.4+5654, and WKK~4438.  
Emission line features are found in all the sources around 6.4~keV in rest frame energy. 

\begin{figure*}
 \begin{center}
  \includegraphics[width=15cm]{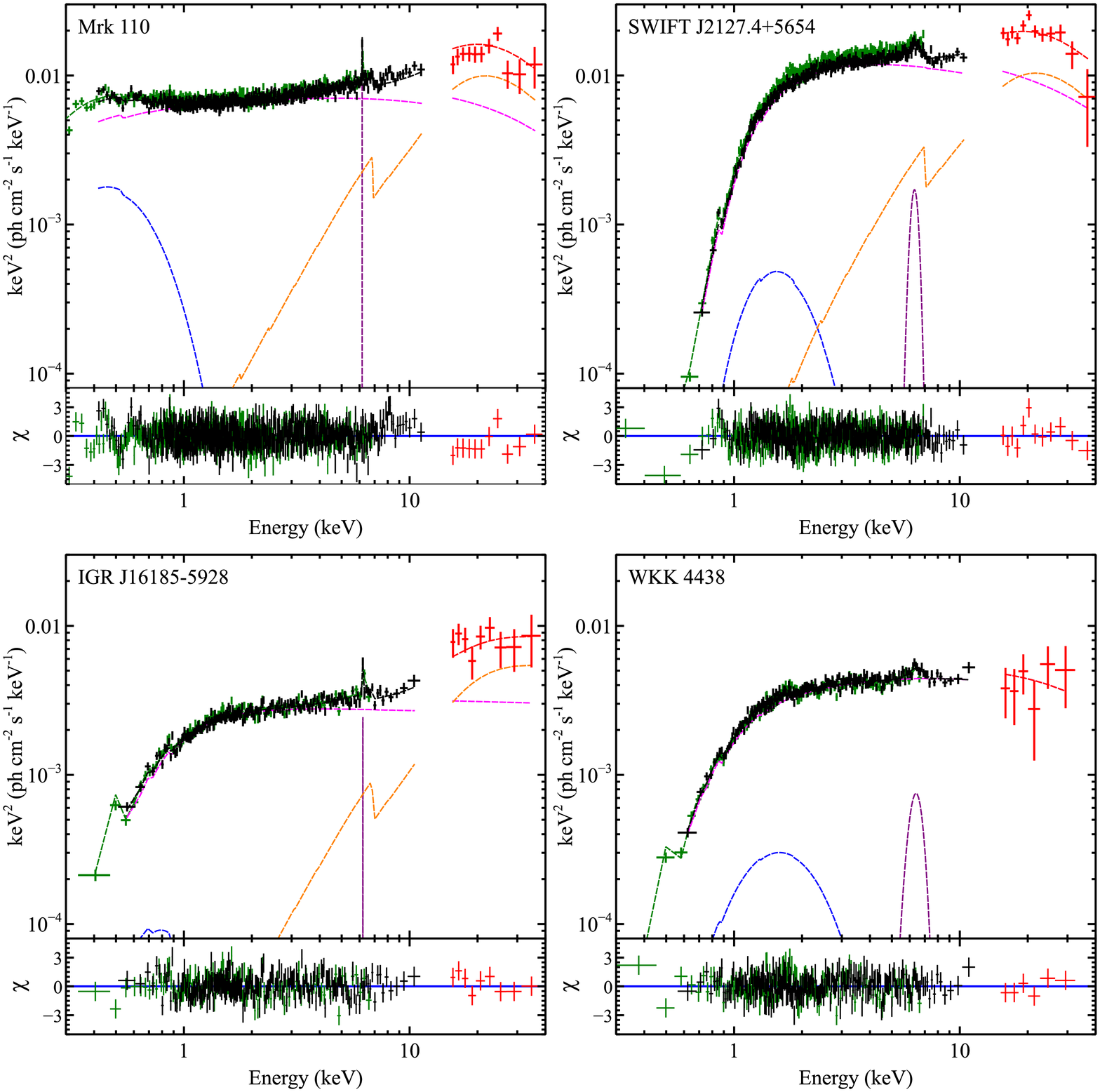} 
 \end{center}
\caption{Results of the spectral fits of Mrk~110, SWIFT~J2127.4+5654, IGR~J16185$-$5928, and WKK~4438 with model A. The crosses are the same as figure~\ref{ratio}, and the dashed lines represent the models. The magenta, orange, purple, and blue lines correspond to the cutoff power law, reflection on cold matter, Fe emission line, and soft X-ray excess component, respectively.}\label{spec_pexrav}
\end{figure*}

The detection of the emission lines at $\sim 6.4~\mathrm{keV}$ provides evidence of reflection on cold matter such as the dusty torus.  
We, therefore, applied the {\tt pexrav} model introduced by \citet{magdziarz1995}, which represents an exponentially cutoff power law component and its reflection continuum. 
The emission line from cold matter was separately described by a Gaussian model ({\tt zgauss}).  
Hereafter, the inclination angle was fixed to $60^{\circ}$ and the abundance of elements in accreting matter was assumed to be solar. 
The normalization of the reflection component is represented by the reflection fraction of $R \equiv \Omega/(2\pi)$, where $\Omega$ is the solid angle of the reflector viewed from the X-ray source.
Since figure~\ref{ratio} shows that Mrk~110 and IGR~J16185$-$5928 have significant soft excess components, a blackbody model ({\tt zbbody}) was added to phenomenologically represent the structures. 
Although SWIFT~J2127.4+5654 and WKK~4438 do not display particularly strong soft excess emission, the fits were improved by including the blackbody model compared with the fits without the blackbody model with $\Delta\chi^2$ of 13.2 and 29.3 for SWIFT~J2127.4+5654 and WKK~4438. 
They correspond to the F-test significance levels of $\sim 3\sigma$ and $\sim 4\sigma$, respectively. 
Thus, we added {\tt zbbody} to the model for SWIFT~J2127.4+5654 and WKK~4438 as well as the other two sources. 
We also checked the presence of the intrinsic absorption in the sources by adding the model of the photoelectric absorption {\tt zphabs} and found no significant additional absorption with the upper limits of $4 \times 10^{18}$, $6 \times 10^{20}$, $8 \times 10^{20}$ and $4 \times 10^{20}~\mathrm{cm}^{-2}$ for Mrk~110, SWIFT~J2127.4+5654, IGR~J16185$-$5928, and WKK~4438, respectively. 
We show the results of the spectral fits with the model {\tt phabs*(pexrav+zgauss+zbbody)} (model A) in figure~\ref{spec_pexrav} and summarize the best-fit parameters in table~\ref{tab4}. 
We confirmed that the line centroids of the emission line are consistent with that of the neutral Fe K$\alpha$ line at 6.4~keV.
The cutoff energies $E_{\mathrm{c}}$ are $\sim 40~\mathrm{keV}$ for Mrk~110, SWIFT~J2127.4+5654, and WKK~4438. 
The lower limit of $E_{\mathrm{c}} > 155~\mathrm{keV}$ was obtained for IGR~J16185$-$5928. 

\begin{table*}
	\begin{center}
	\tbl{Best-fit parameters with model A.}{%
  \begin{tabular}{cccccc}\hline
		Model & Parameter & Mrk~110 & SWIFT~J2127.4+5654 & IGR~J16185$-$5928 & WKK~4438 \\ \hline
		{\tt pexrav} & $\Gamma$ & $1.84 \pm 0.02$ & $2.01 \pm 0.05$  & $2.04 \pm 0.03$ & $1.79^{+0.05}_{-0.06}$\\ 
		& $E_{\mathrm{c}}$ (keV) & $35^{+10}_{-6}$ & $38^{+23}_{-9}$ & $> 155$ & $35^{+30}_{-16}$ \\  
		& Normalization\footnotemark[$*$] & $(6.3 \pm 0.1)\times 10^{-3}$ & $(1.3 \pm 0.1)\times 10^{-2}$  & $(3.0 \pm 0.1)\times 10^{-3}$ & $(3.5 \pm 0.1)\times 10^{-3}$  \\ 
		& $R$ & $2.7 \pm 0.6$ & $2.2 \pm 0.9$ & $2.6^{+0.7}_{-0.6}$ & $< 3.3$ \\ \hline
		{\tt zgauss} & Line energy (keV) & $6.38 \pm 0.03$ & $6.37 \pm 0.09$ & $6.43^{+0.04}_{-0.03}$ & $6.43 \pm 0.14$ \\ 
		& Line width (keV) & $< 0.08$ & $0.26^{+0.19}_{-0.13}$ & $< 0.08$ & $< 0.66$ \\
		& Normalization\footnotemark[$*$] & $(0.9 \pm 0.3)\times 10^{-5}$ & $(2.8^{+2.0}_{-1.3})\times 10^{-5}$ & $(6.0^{+2.1}_{-1.8})\times 10^{-6}$ & $(2.2 \pm 0.9)\times 10^{-5}$ \\
		& Equivalent width (eV) & $35 \pm 10$ & $77^{+39}_{-35}$ & $64^{+24}_{-25}$ & $192 \pm 45$ \\ \hline
		{\tt zbbody} & $kT_{\mathrm{bb}}$ (keV) & $0.11 \pm 0.01$ & $0.27^{+0.03}_{-0.02}$ & $0.13^{+0.03}_{-0.02}$ & $0.35^{+0.05}_{-0.04}$ \\
		& Normalization\footnotemark[$*$] & $(5.7 \pm 0.5)\times 10^{-5}$ & $(3.4 \pm 1.5)\times 10^{-5}$ & $(1.1 \pm 0.6)\times 10^{-5}$ & $(1.1^{+0.4}_{-0.3})\times 10^{-5}$ \\ \hline
		& $\chi^2$/dof & $1134.85/972$ & $735.19/717$ & $379.43/329$ & $446.74/375$ \\ \hline
	\end{tabular}}\label{tab4}
    \end{center}
\begin{tabnote}
	\footnotemark[$*$]  in units of $\mathrm{ph~keV^{-1}~cm^{-2}~s^{-1}}$. \\
\end{tabnote}
\end{table*}

\begin{figure*}
 \begin{center}
  \includegraphics[width=15cm]{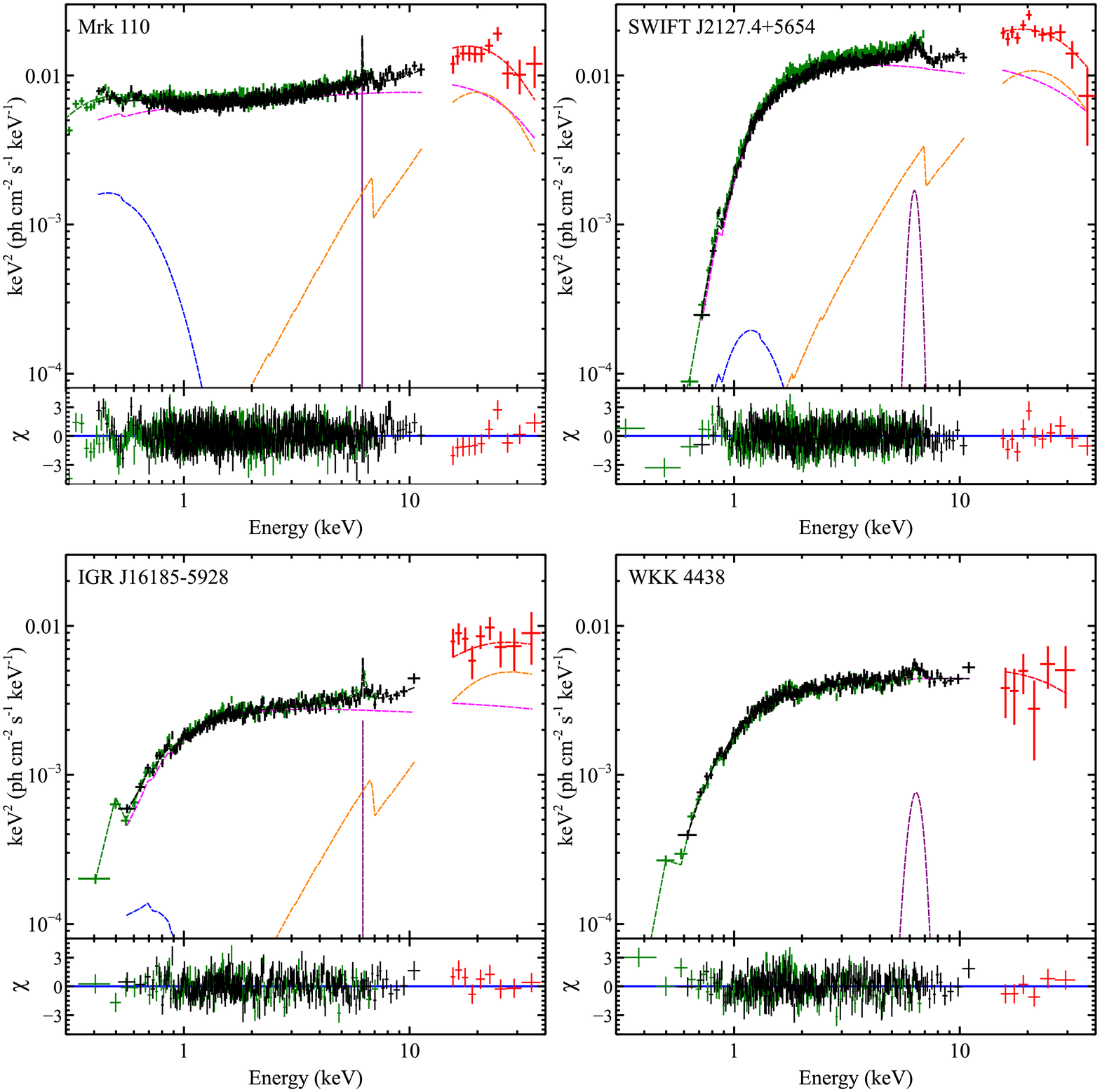} 
 \end{center}
\caption{Results of the spectral fits of Mrk~110, SWIFT~J2127.4+5654, IGR~J16185$-$5928, and WKK~4438 with model B. The crosses and lines are the same as figure~\ref{spec_pexrav} but the magenta lines represent the thermal Comptonization component.}\label{spec_nthcomp}
\end{figure*}

\begin{table*}
	\begin{center}
	\tbl{Best-fit parameters with model B.}{%
  \begin{tabular}{cccccc}\hline
		Model & Parameter & Mrk~110 & SWIFT~J2127.4+5654 & IGR~J16185$-$5928 & WKK~4438 \\ \hline
		{\tt nthcomp} & $\Gamma$ & $1.92 \pm 0.02$ & $2.13 \pm 0.04$ & $2.05^{+0.03}_{-0.04}$ & $1.95^{+0.02}_{-0.01}$\\ 
		& $kT_{\mathrm{e}}$ (keV) & $7.4^{+0.8}_{-0.7}$ & $11^{+5}_{-2}$ & $> 18$ & $> 5$ \\  
		& Normalization\footnotemark[$*$] & $(6.4 \pm 0.1)\times 10^{-3}$& $(1.4 \pm 0.1)\times 10^{-2}$ & $(3.0 \pm 0.1)\times 10^{-3}$ & $(3.9 \pm 0.1)\times 10^{-3}$\\ \hline
		{\tt reflect} & $R$ & $2.1 \pm 0.4$ & $2.3^{+1.0}_{-0.8}$ & $2.7^{+0.7}_{-0.8}$ & $< 0.4$ \\ \hline
		{\tt zgauss} & Line energy (keV) & $6.38 \pm 0.03$ & $6.36^{+0.09}_{-0.11}$ & $6.42 \pm 0.04$ & $6.43 \pm 0.14$  \\ 
		& Line width (keV) & $< 0.08$ & $0.3 \pm 0.2$ & $< 0.09$ & $0.5 \pm 0.2$  \\
		& Normalization\footnotemark[$*$] & $(0.9 \pm 0.3)\times 10^{-5}$ & $(3.5^{+2.2}_{-1.7})\times 10^{-5}$ & $< 6.5 \times 10^{-6}$& $(2.2^{+0.8}_{-0.7})\times 10^{-5}$\\
		& Equivalent width (eV) & $36 \pm 10$  & $94^{+40}_{-38}$ & $< 61$ & $199^{+204}_{-81}$\\ \hline
		{\tt zbbody} & $kT_{\mathrm{bb}}$ (keV) & $0.099 \pm 0.003$ & $0.16^{+0.02}_{-0.04}$ & $0.11 \pm 0.01$ & $0.14 \pm 0.01$ \\
		& Normalization\footnotemark[$*$] & $(8.2 \pm 0.53)\times 10^{-5}$ & $< 7.6 \times 10^{-5}$ & $(2.3 \pm 0.7)\times 10^{-5}$& $< 0.1\times 10^{-5}$\\ \hline
		& $F_{0.3-40~\mathrm{keV}}~(\mathrm{erg~cm^{-2}~s^{-1}})$\footnotemark[$\dagger$] & $8.9 \times 10^{-11}$ & $8.7 \times 10^{-11}$ & $2.9 \times 10^{-11}$ & $2.8 \times 10^{-11}$ \\
		& $L_{0.3-40~\mathrm{keV}}~(\mathrm{erg~s^{-1}})$\footnotemark[$\dagger$] & $2.4 \times 10^{44}$ & $4.1 \times 10^{43}$ & $8.0 \times 10^{43}$ & $1.5 \times 10^{43}$ \\ 
		& $F_{\mathrm{Comp}}~(\mathrm{erg~cm^{-2}~s^{-1}})$\footnotemark[$\ddagger$] & $5.3 \times 10^{-11}$ & $8.8 \times 10^{-11}$ & $>2.3 \times 10^{-11}$ & $>2.7 \times 10^{-11}$ \\
		& $L_{\mathrm{Comp}}~(\mathrm{erg~s^{-1}})$\footnotemark[$\ddagger$] & $1.4 \times 10^{44}$ & $4.1 \times 10^{43}$ & $>6.3 \times 10^{43}$ & $>1.5 \times 10^{43}$ \\ \hline 
		& $\chi^2$/dof & $1129.27/972$ & $735.74/717$ & $358.94/329$ & $458.36/375$\\ \hline
    \end{tabular}}\label{tab5}
    \end{center}
\begin{tabnote}
	\footnotemark[$*$]  in units of $\mathrm{ph~keV^{-1}~cm^{-2}~s^{-1}}$. \\
	\footnotemark[$\dagger$] absorbed flux or luminosity in 0.3--40~keV. \\
	\footnotemark[$\ddagger$] unabsorbed flux or luminosity of the Comptonized component. \\
\end{tabnote}
\end{table*}

The continuum described by a cutoff power law can be ascribed to Compton up-scattering of seed photons by accreting electrons \citep{haardt1994}. 
The cutoff energy is related to the electron temperature $kT_{\mathrm{e}}$ as $E_{\mathrm{c}} \approx 2$--$3\, kT_{\mathrm{e}}$ \citep{petrucci2001}. 
In order to directly obtain $kT_{\mathrm{e}}$ from spectral fittings, we used the thermal Comptonization model ({\tt nthcomp}; \cite{zdziarski1996}; \cite{zycki1999}) instead of the {\tt pexrav} model, and the convolution model for reflection on cold matter ({\tt reflect}; \cite{magdziarz1995}). 
The seed photon temperature $kT_{\mathrm{s}}$ was assumed to be the same as the blackbody temperature $kT_{\mathrm{bb}}$ of the {\tt zbbody} model. 
The results of the spectral fits with the model {\tt phabs*[(1+reflect)*nthcomp+zgauss+zbbody]} (model B) are shown in figure~\ref{spec_nthcomp} and the best-fit parameters are in table~\ref{tab5}. 
While we assumed the source of the Compton seed photons, namely the soft excess, to be a blackbody, some alternative interpretations are proposed by several authors as well. 
For example, some claimed that the soft excess can be explained as the hardest end of a Comptonized accretion disk emission which is originated from an additional cool and thick corona (e.g., Done et al.\ \yearcite{done2012}, \yearcite{done2013}; \cite{jin2012}; \cite{petrucci2013}). 
If this is the case, $kT_{\mathrm{s}}$ could be overestimated/underestimated in our analysis. 
Thus, the assumption on $kT_{\mathrm{s}}$ is rather model dependent. 
Therefore, we also fit the spectra allowing $kT_{\mathrm{s}}$ to vary as a free parameter independent of $kT_{\mathrm{bb}}$. 
The upper limit or best-fit values for $kT_{\mathrm{s}}$ were $< 0.06$, $0.15^{+0.09}_{-0.02}$, $0.20^{+0.03}_{-0.12}$, and $0.14 \pm 0.01~\mathrm{keV}$ for Mrk~110, SWIFT~J2127.4+5654, IGR~J16185$-$5928, and WKK~4438, respectively. 
They are equivalent with (SWIFT~J2127.4+5654, IGR~J16185$-$5928, and WKK~4438) or slightly lower (Mrk~110) than $kT_{\mathrm{bb}}$. 
In addition, we found that the fits do not significantly change the other parameters given in table~\ref{tab5}, in particular $kT_{\mathrm{e}}$ and $\Gamma$ which are the key parameters for the main subject of this paper.
Hence, we hereafter show only the results obtained under the assumption that $kT_{\mathrm{s}}$ is equivalent to $kT_{\mathrm{bb}}$ for simplicity. 




In our sample, WKK~4438 has a relatively weak reflection continuum with the reflection fractions of $R < 0.4$. 
On the other hand, Mrk~110, SWIFT~J2127.4+5654, and IGR~J16185$-$5928 exhibit a high reflection fractions of $R \sim 2$. 
The high $R$ value suggests possible contributions from additional reflection components. 
Since Fe K$\alpha$ emission lines in Mrk~110 and IGR~J16185$-$5928 are narrow, the additional component should be cold similarly to the dusty torus and, therefore, would be attributed to the outer part of the accretion disk.
In contrast to the two sources, we detected a broad Fe K$\alpha$ emission line in the SWIFT~J2127.4+5654 spectrum, as already reported by \citet{miniutti2009} and \citet{patrick2011}, which provides firm evidence of the reflection component from the inner part of the accretion disk. 
This is because the broad emission line is considered to be due to relativistic effects such as Doppler motions and gravitational redshifts \citep{fabian1989} and/or many unresolved lines from various ionization states \citep{ross1993}. 

\begin{figure}
 \begin{center}
  \includegraphics[bb=90 0 350 400, width=5cm]{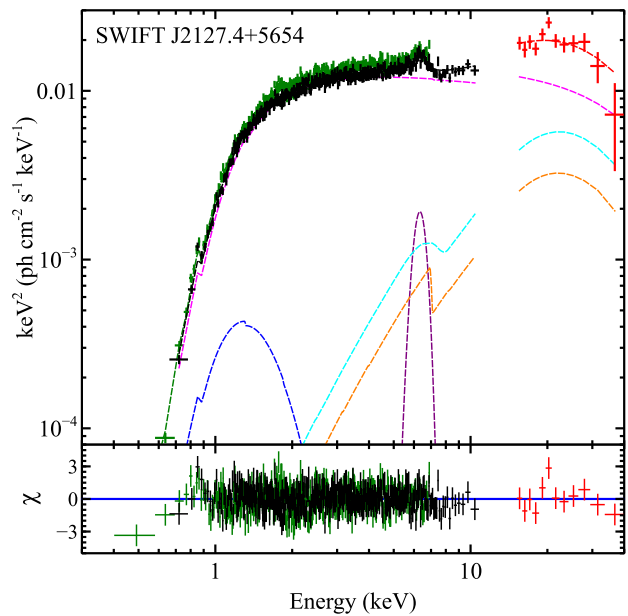} 
 \end{center}
\caption{Result of the spectral fit of SWIFT~J2127.4+5654 with model C. The crosses and lines are the same as figure~\ref{spec_nthcomp} and the cyan line represents the reflection component on the accretion disk.}\label{spec_bluriref}
\end{figure}

\begin{table}
	\begin{center}
	\tbl{Best-fit parameters with model C.}{%
  \begin{tabular}{ccc}\hline
		Model & Parameter & SWIFT~J2127.4+5654 \\ \hline
		{\tt nthcomp}& $\Gamma$ & $2.08^{+0.05}_{-0.04}$ \\ 
		& $kT_{\mathrm{e}}$ (keV) & $12^{+7}_{-3}$ \\ 
		& Normalization\footnotemark[$*$] & $(1.2^{+0.2}_{-0.1})\times 10^{-2}$\\ \hline
		{\tt reflect} & $R_{\mathrm{cold}}$ & $0.59$\footnotemark[$\dagger$] \\ \hline
		{\tt ireflect} & $\xi~(\mathrm{erg~cm~s^{-1}})$ & $< 18$\\
		& $R_{\mathrm{disk}}$ & $1.0^{+0.9}_{-0.7}$ \\ \hline
		{\tt zgauss} & Line energy (keV) & $6.38 \pm 0.10$ \\ 
		& Line width (keV) & $0.4 \pm 0.2$ \\
		& Normalization\footnotemark[$*$] & $(4.6^{+1.7}_{-1.4})\times 10^{-5}$\\
		& Equivalent width (eV) & $129^{+40}_{-45}$ \\ \hline
		{\tt zbbody} & $kT_{\mathrm{bb}}$ (keV) & $0.20 \pm 0.02$ \\
		& Normalization\footnotemark[$*$] & $(5.5^{+2.0}_{-2.7})\times 10^{-5}$\\ \hline
		& $F_{0.3-40~\mathrm{keV}}~(\mathrm{erg~cm^{-2}~s^{-1}})$\footnotemark[$\ddagger$] & $8.6 \times 10^{-11}$  \\
		& $L_{0.3-40~\mathrm{keV}}~(\mathrm{erg~s^{-1}})$\footnotemark[$\ddagger$] & $4.0 \times 10^{43}$  \\ 
		& $F_{\mathrm{Comp}}~(\mathrm{erg~cm^{-2}~s^{-1}})$\footnotemark[$\S$] & $8.6 \times 10^{-11}$  \\
		& $L_{\mathrm{Comp}}~(\mathrm{erg~s^{-1}})$\footnotemark[$\S$] & $4.0 \times 10^{43}$ \\ \hline
		& $\chi^2$/dof & $728.88/716$ \\ \hline
    \end{tabular}}\label{bluriref}
    \end{center}
\begin{tabnote}
	\footnotemark[$*$]  in units of $\mathrm{ph~keV^{-1}~cm^{-2}~s^{-1}}$. \\
	\footnotemark[$\dagger$]  fixed value. \\
	\footnotemark[$\ddagger$]  absorbed flux or luminosity in 0.3--40~keV. \\
	\footnotemark[$\S$] unabsorbed flux or luminosity of the Comptonized component. \\ 
\end{tabnote}
\end{table} 

We added to the SWIFT~J2127.4+5654 spectrum a reflection continuum model in which a thermal Comptonization model ({\tt nthcomp}) is convolved with the model simulating the relativistic effects ({\tt kdblur}; \cite{laor1991}) and also with the ionized reflection model ({\tt ireflect}; \cite{magdziarz1995}).  
The outer disk radius was assumed to be $400r_{\mathrm{g}}$, the value \citet{patrick2011} fixed to, where $r_{\mathrm{g}}$ is the gravitational radius. 
In accordance with the result by \citet{patrick2011}, the inner disk radius and the radial emissivity index of the disk were fixed to $20r_{\mathrm{g}}$ and $2.5$, respectively. 
The disk temperature of the {\tt ireflect} model was assumed to be $10^{5}~\mathrm{K}$. 
The ionization parameter of the disk is defined as $\xi = 4\pi F_{\mathrm{irr}}/n_{\mathrm{disk}}$, where $F_{\mathrm{irr}}$ is the 5~eV--20~keV flux of photons irradiating the disk and $n_{\mathrm{disk}}$ is the gas density of the disk. 
Since the fraction of the reflection on the accretion disk $R_{\mathrm{disk}}$ and on cold matter $R_{\mathrm{cold}}$ are mutually dependent, they were not determined uniquely when both parameters were allowed to vary. Hence, $R_{\mathrm{cold}}$ was fixed to 0.59, the average value of lightly obscured AGNs obtained by \citet{kawamuro2016}. We show the result of the spectral fit with the model {\tt phabs*[(1+reflect+kdblur*ireflect)*nthcomp+zgauss\\{}+zbbody]} (model C) in figure~\ref{spec_bluriref} and the best-fit parameters in table~\ref{bluriref}. 
The upper limit of $\xi$ is $18~\mathrm{erg~cm~s^{-1}}$, which indicates that the ionized reflection is hardly required. 

We calculate the optical depth of the corona from the results with model B for Mrk~110, IGR~J16185$-$5928, and WKK~4438, and with model C for SWIFT~J2127.4+5654. 
The parameter $\Gamma$ is related to the Compton $y$ parameter with the following equation \citep{rybicki1979}:
\begin{equation}
	(\Gamma - 1)(\Gamma + 2)-\frac{4}{y} = 0.
\end{equation}
When the corona is optically thick, the $y$ parameter is approximated as 
\begin{equation}
	y = \frac{4kT_{\mathrm{e}}}{m_{\mathrm{e}}c^2}\mathrm{max}(\tau,\tau^2) \sim \frac{4kT_{\mathrm{e}}}{m_{\mathrm{e}}c^2}\tau\left(1+\frac{\tau}{3}\right)
\end{equation}
in the thermal Comptonization \citep{petrucci2001}.
Thus, the optical depth $\tau$ is approximated with the following equation \citep{sunyaev1980,lightman1987}:
\begin{equation}\label{tau}
	\tau=\sqrt{\mathstrut \frac{9}{4}+\frac{3}{(kT_{\mathrm{e}}/m_{\mathrm{e}}c^2)(\Gamma-1)(\Gamma+2)}}-\frac{3}{2}.
\end{equation}
The $y$ parameter and $\tau$ are listed in table~\ref{tab6}. 
The corona is found to be optically thick at least for Mrk~110 and SWIFT~J2127.4+5654. 

\begin{table*}
	\begin{center}
	\tbl{Compton $y$ parameter and $\tau$.}{%
  \begin{tabular}{ccccc}\hline
		Parameter & Mrk~110 & SWIFT~J2127.4+5654 & IGR~J16185$-$5928 & WKK~4438 \\ \hline
		$y$ & $1.11\pm 0.03$ & $0.91^{+0.05}_{-0.06}$ & $0.94^{+0.05}_{-0.03}$ & $1.06^{+0.02}_{-0.03}$ \\
		$\tau$ & $6.3^{+0.4}_{-0.5}$ & $4.1^{+0.8}_{-1.5}$ & $< 3.2$ & $< 7.3$ \\ \hline
    \end{tabular}}\label{tab6}
    \end{center}
\begin{tabnote}
\end{tabnote}
\end{table*}

\subsection{NuSTAR Analysis} 
We also analyzed the NuSTAR data of SWIFT~J2127.4+5654 (table~\ref{obs}). 
The same data were already analyzed by \citet{marinucci2014}. 
They performed spectral fittings with a common value of $kT_{\mathrm{e}}$ regardless of the change in luminosity. 
On the other hand, as described below, the purpose of our analysis is to study the correlation between $kT_{\mathrm{e}}$ and the luminosity. 

\begin{figure}
 \begin{center}
  \includegraphics[bb=50 0 500 390, width=7cm]{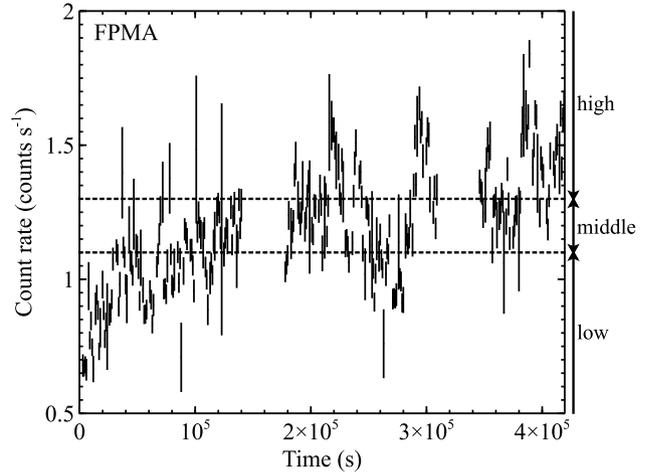} 
 \end{center}
\caption{Background inclusive FPMA lightcurve of SWIFT~J2127.4+5654 in 3--79~keV. The x-axis represents the duration from the beginning of the observation. Each data bin has 1~ks in width.}\label{curve}
\end{figure}

\begin{figure}
 \begin{center}
  \includegraphics[bb=90 0 350 400, width=5cm]{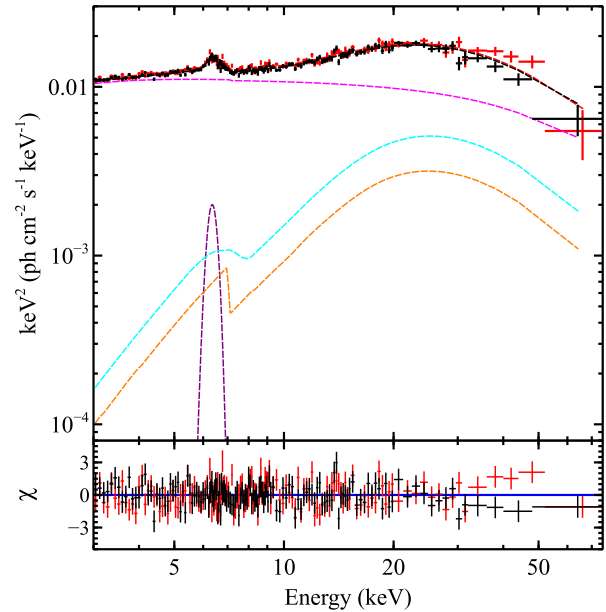} 
 \end{center}
\caption{Time averaged spectra of SWIFT~J2127.4+5654. The black and red crosses represent FPMA and FPMB, respectively. The lines are the same as figure~\ref{spec_bluriref}.}\label{spec_ave}
\end{figure}

Figure~\ref{curve} shows the FPMA lightcurve including background. The count rate of background was stable at about $0.04~\mathrm{counts~s^{-1}}$. In this observation, the luminosity changed about three times from peak to peak. 
We first fit time averaged spectra in the 3--78~keV band with the same model as model C but without the {\tt zbbody} model, {\tt phabs*[(1+reflect+kdblur*ireflect)*nthcomp+zgauss]} (see figure~\ref{spec_ave}), since the soft excess is negligible above 3~keV. 
The seed photon temperature was fixed to $kT_{\mathrm{bb}}=0.1~\mathrm{keV}$. 
The other parameters were fixed to the same values as used in the Suzaku analysis. 
The cross-normalization factor between FPMA and FPMB was allowed to vary. The best-fit parameters are shown in table~\ref{nuave}. 
The ionization parameter is again constrained to be very small ($\xi < 22~\mathrm{erg~cm~s^{-1}}$), being consistent with the Suzaku result in section \ref{suzakuana}. 
The fit gave an electron temperature of $kT_{\mathrm{e}} = 20^{+5}_{-2}~\mathrm{keV}$. 
It is considerably lower than that obtained with the same NuSTAR data by \citet{marinucci2014}. 
We consider this is due to the difference in CALDB used between the two analyses. 
A major update of CALDB of NuSTAR was made in 2016 June \footnote{$\langle$https://heasarc.gsfc.nasa.gov/docs/heasarc/caldb/nustar/docs/release\_\\{}20160606.txt$\rangle$}, in which the low energy effective area of the instrument was revised based on an observation of Crab at a large off-axis angle \citep{madsen2017}.

\begin{table}
	\begin{center}
	\tbl{Best-fit parameters of the time averaged spectra.}{%
  \begin{tabular}{ccc}\hline
		Model & Parameter & time average \\ \hline
		{\tt nthcomp}& $\Gamma$ & $2.05 \pm 0.03$ \\ 
		& $kT_{\mathrm{e}}$ (keV) & $20^{+5}_{-2}$ \\ 
		& Normalization\footnotemark[$*$] & $(1.24 \pm 0.04)\times 10^{-2}$\\ \hline
		{\tt reflect} & $R_{\mathrm{cold}}$ & $0.59$\footnotemark[$\dagger$] \\ \hline
		{\tt ireflect}& $\xi~(\mathrm{erg~cm~s^{-1}})$ & $< 22$\\
		& $R_{\mathrm{disk}}$ & $1.0^{+0.2}_{-0.3}$ \\ \hline
		{\tt zgauss}& Line energy (keV) & $6.43 \pm 0.05$ \\ 
		& Line width (keV) & $0.2 \pm 0.1$ \\
		& Normalization\footnotemark[$*$] & $(2.8 \pm 0.4)\times 10^{-5}$\\
		& Equivalent width (eV) & $86^{+17}_{-11}$ \\ \hline
		& $\chi^2$/dof & $324.85/293$ \\ \hline
    \end{tabular}}\label{nuave}
    \end{center}
\begin{tabnote}
	\footnotemark[$*$]  in units of $\mathrm{ph~keV^{-1}~cm^{-2}~s^{-1}}$. \\
	\footnotemark[$\dagger$]  fixed value. \\
\end{tabnote}
\end{table}

\begin{figure}
 \begin{center}
  \includegraphics[bb=90 0 350 560, width=5cm]{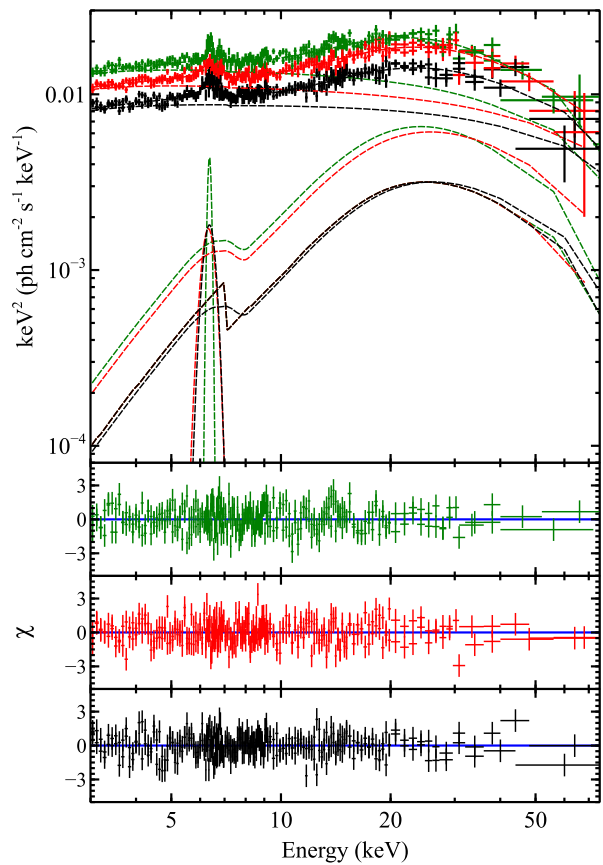} 
 \end{center}
\caption{Spectra of SWIFT~J2127.4+5654 in the high (green crosses), middle (red crosses), and low (black crosses) flux period. The dashed lines represent models.}\label{spec_hml}
\end{figure}

\begin{figure}
 \begin{center}
  \includegraphics[bb=90 0 350 400, width=5cm]{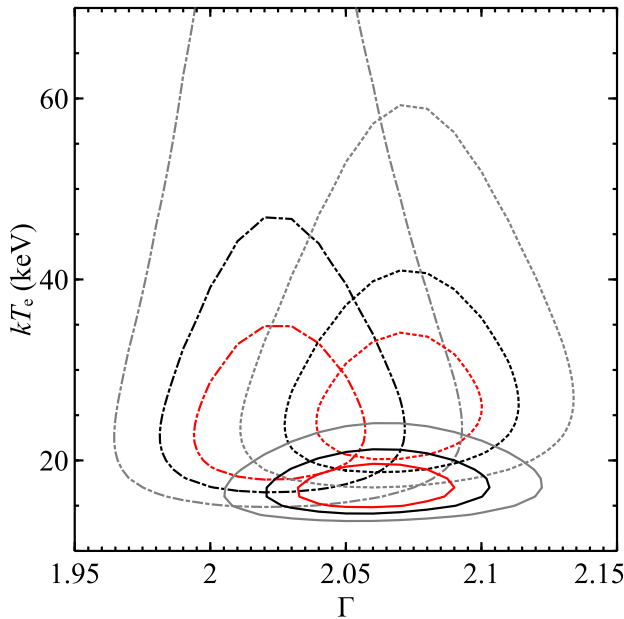} 
 \end{center}
\caption{Confidence contours of $kT_{\mathrm{e}}$ versus $\Gamma$ in NuSTAR results of SWIFT~J2127.4+5654. The solid, dotted, and dot-dashed lines correspond to the high, middle, and low flux period, and red, black, and gray lines represent $1\sigma$, $90\%$, and $99\%$ confidence level, respectively.}\label{cont}
\end{figure}

In order to search for possible spectral changes, we divided the FPMA and FPMB datasets according to the count rate of FPMA into three, the high, middle, and low flux periods, as shown in figure~\ref{curve}. 
We fit the spectra of the three periods with the same model as that used for the time averaged spectra. 
We can safely assume that the reflection component on the cold matter is stable during the observation (5 days), considering the torus size (e.g., $\sim 300$~light-days; \cite{kynoch2018}) and its viewing solid angle ($\Omega = 1.18\pi$ when $R_{\mathrm{cold}} = 0.59$). 
We used the model {\tt phabs*[reflect*nthcomp$_1$+(1+kdblur*ireflect)*nthcomp$_2$\\{}+zgauss]}, in which the parameters for the reflection component on the cold matter ({\tt reflect*nthcomp$_1$}) were fixed to the best-fit values from the time averaged spectrum fit (table~\ref{nuhml}). 
We also fixed the ionization parameter for the reflection component on the inner disk ({\tt kdblur*ireflect*nthcomp$_2$}) to $\xi = 0$, since the contribution of the ionized reflection is found to be negligible with the time averaged spectrum.  
We show the results of the spectral fit in figure~\ref{spec_hml}.
In the high flux period, $kT_{\mathrm{e}}$ is slightly lower than those in the middle and low flux periods.
Figure~\ref{cont} shows the confidence contours of the mutually dependent parameters $\Gamma$ and $kT_{\mathrm{e}}$, suggesting marginal difference of the spectral shape among the three periods.  

\begin{table*}
	\begin{center}
	\tbl{Best-fit parameters of the spectra in each flux period.}{%
  \begin{tabular}{ccccc}\hline
		Model & Parameter & high & middle & low \\ \hline
		{\tt nthcomp$_1$}\footnotemark[$\ddagger$] & $\Gamma$ & $2.05$\footnotemark[$\dagger$] & $2.05$\footnotemark[$\dagger$] & $2.05$\footnotemark[$\dagger$] \\ 
		& $kT_{\mathrm{e}}$ (keV) & $20$\footnotemark[$\dagger$] & $20$\footnotemark[$\dagger$] & $20$\footnotemark[$\dagger$] \\ 
		& Normalization\footnotemark[$*$] & $1.24 \times 10^{-2}$\footnotemark[$\dagger$] & $1.24 \times 10^{-2}$\footnotemark[$\dagger$] & $1.24 \times 10^{-2}$\footnotemark[$\dagger$] \\ \hline
		{\tt reflect}\footnotemark[$\ddagger$] & $R_{\mathrm{cold}}$ & $0.59$\footnotemark[$\dagger$] & $0.59$\footnotemark[$\dagger$] & $0.59$\footnotemark[$\dagger$] \\ \hline
		{\tt nthcomp$_2$}& $\Gamma$ & $2.06 \pm 0.04$ & $2.07 \pm 0.04$ & $2.02^{+0.04}_{-0.03}$ \\ 
		& $kT_{\mathrm{e}}$ (keV) & $17 \pm 3$ & $25^{+11}_{-6}$ & $23^{+15}_{-6}$ \\ 
		& Normalization\footnotemark[$*$] & $(1.56^{+0.08}_{-0.07})\times 10^{-2}$ & $(1.29^{+0.06}_{-0.07})\times 10^{-2}$ & $(0.93 \pm 0.05)\times 10^{-2}$ \\ \hline
		{\tt ireflect} & $\xi~(\mathrm{erg~cm~s^{-1}})$ & $0$\footnotemark[$\dagger$] & $0$\footnotemark[$\dagger$] & $0$\footnotemark[$\dagger$] \\
		& $R_{\mathrm{disk}}$ & $1.1 \pm 0.3$ & $1.1^{+0.4}_{-0.3}$ & $0.7 \pm 0.3$ \\ \hline
		{\tt zgauss} & Line energy (keV) & $6.46^{+0.07}_{-0.08}$ & $6.39 \pm 0.10$ & $6.43^{+0.09}_{-0.08}$ \\ 
		& Line width (keV) & $< 0.3$ & $0.3^{+0.1}_{-0.2}$ & $0.3^{+0.1}_{-0.2}$ \\
		& Normalization\footnotemark[$*$] & $(2.2^{+0.9}_{-0.6})\times 10^{-5}$ & $(3.0^{+0.9}_{-0.8})\times 10^{-5}$ & $(2.9^{+0.8}_{-0.7})\times 10^{-5}$ \\
		& Equivalent width (eV) & $56^{+12}_{-16}$ & $88^{+27}_{-17}$ & $113^{+25}_{-21}$ \\ \hline
		& $F_{3-78~\mathrm{keV}}~(\mathrm{erg~cm^{-2}~s^{-1}})$\footnotemark[$\S$] & $8.1 \times 10^{-11}$ & $7.2 \times 10^{-11}$ & $5.5 \times 10^{-11}$  \\
		& $L_{3-78~\mathrm{keV}}~(\mathrm{erg~s^{-1}})$\footnotemark[$\S$] & $3.8 \times 10^{43}$ & $3.4 \times 10^{43}$ & $2.6 \times 10^{43}$  \\ 
		& $F_{\mathrm{Comp}}~(\mathrm{erg~cm^{-2}~s^{-1}})$\footnotemark[$\|$] & $1.2 \times 10^{-10}$ & $1.0 \times 10^{-10}$ & $7.9 \times 10^{-11}$  \\
		& $L_{\mathrm{Comp}}~(\mathrm{erg~s^{-1}})$\footnotemark[$\|$] & $5.6 \times 10^{43}$ & $4.7 \times 10^{43}$ & $3.7 \times 10^{43}$  \\ \hline
		& $\chi^2$/dof & $322.20/294$ & $302.35/294$ & $261.12/294$ \\ \hline
    \end{tabular}}\label{nuhml}
    \end{center}
\begin{tabnote}
	\footnotemark[$*$]  in units of $\mathrm{ph~keV^{-1}~cm^{-2}~s^{-1}}$. \\
	\footnotemark[$\dagger$]  fixed value. \\
	\footnotemark[$\ddagger$]  These models represent the stable reflection component from the distant cold matter ({\tt reflect*nthcomp$_1$}). \\
	\footnotemark[$\S$] absorbed flux or luminosity in 3--78~keV. \\
	\footnotemark[$\|$] unabsorbed flux or luminosity of the Comptonized component. \\
\end{tabnote}
\end{table*}

\section{Discussion}

\subsection{Comparison of the Nature of the Corona between NLS1s and BLS1s}\label{cutoff}
We studied the X-ray wide band spectra of the four NLS1s.  
$E_{\mathrm{c}}$ of Mrk~110 is successfully constrained in this paper for the first time.
$E_{\mathrm{c}}$ of SWIFT~J2127.4+5654, IGR~J16185$-$5928, and WKK~4438 obtained in this paper agree with those reported by \citet{malizia2008}, while we constrained the value with smaller error for WKK~4438. 
For SWIFT~J2127.4+5654, $E_{\mathrm{c}}$ in our result is also consistent with those of \citet{panessa2011} and \citet{miniutti2009}.

\begin{figure}
 \begin{center}
  \includegraphics[bb=90 0 350 400, width=5cm]{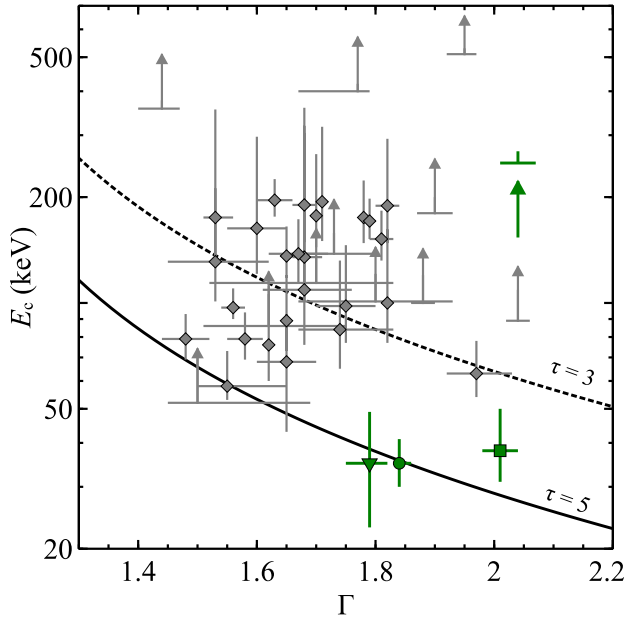} 
 \end{center}
\caption{Scatter plot of $E_{\mathrm{c}}$ and $\Gamma$. The green circle, square and triangle represent Mrk~110, SWIFT~J2127.4+5654 and WKK~4438, respectively. The green arrow shows the lower limit obtained from IGR~J16185$-$5928. The gray diamonds and arrows correspond to BLS1s \citep{malizia2014}. For BLS1s, the 90\% confidence range was divided by 1.645, converted to $1\sigma$. The lower limits are at 90\% confidence levels. The black lines represent equation~(\ref{tau}) with a simple estimation as $E_{\mathrm{c}}=3kT_{\mathrm{e}}$ \citep{petrucci2001}.}\label{ecplot}
\end{figure}

For BLS1s, \citet{malizia2003} reported the average cutoff energy of $E_{\mathrm{c}} = 216^{+75}_{-41}~\mathrm{keV}$. 
The same group reported the mean value of $E_{\mathrm{c}} = 128~\mathrm{keV}$ with a spread of $46~\mathrm{keV}$ \citep{malizia2014}. 
Except for IGR~J16185$-$5928, our spectral analyses of the NLS1s gave cutoff energies of $E_{\mathrm{c}} \sim 40~\mathrm{keV}$, which are lower than those of BLS1s. 
This is consistent with the suggestion by \citet{malizia2008} that NLS1s have lower $E_{\mathrm{c}}$ than BLS1s. 
\citet{malizia2008} pointed out also that NLS1s have steeper spectra than BLS1s. 
In order to systematically study these differences between NLS1s and BLS1s, we present relation between $\Gamma$ and $E_{\mathrm{c}}$ from the {\tt pexrav} model fits (model A) in figure~\ref{ecplot}, together with results on BLS1s by \citet{malizia2014}.  
Our NLS1 sample has typically larger $\Gamma$ and lower $E_{\mathrm{c}}$ than the BLS1 sample, confirming the claim by \citet{malizia2008} that NLS1s have softer spectra and lower $kT_{\mathrm{e}}$. 
The two parameters $\Gamma$ and $E_{\mathrm{c}}$ are statistically coupled and often show a positive correlation. 
The correlation we found, however, is a negative one, which cannot be explained by the statistical coupling. 
We also found that the corona of our NLS1 sample tends to be optically thicker than that of the BLS1 sample in figure~\ref{ecplot} (see also table~\ref{tab6} and \cite{petrucci2001}). 

\subsection{Dependence of the electron temperature on the accretion rate}  
\begin{figure}
 \begin{center}
  \includegraphics[bb=90 0 350 400, width=5cm]{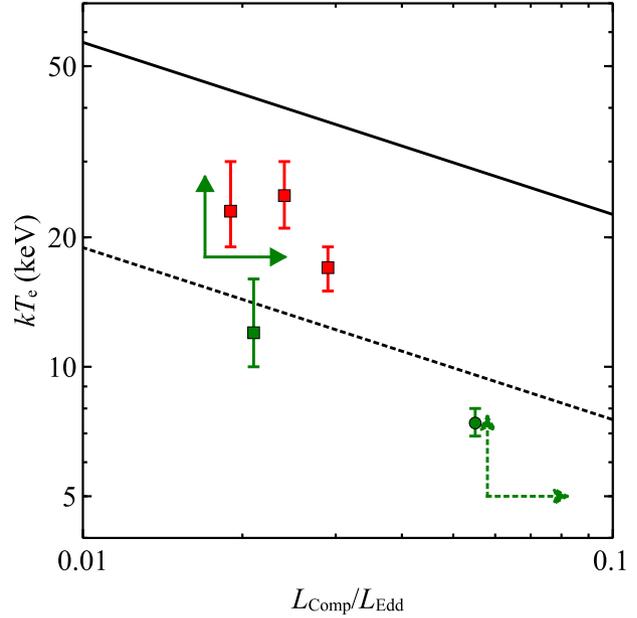} 
 \end{center}
\caption{Plot of $kT_\mathrm{e}$ versus $L_{\mathrm{Comp}}/L_{\mathrm{Edd}}$. Green and red correspond to the Suzaku results and NuSTAR results, respectively. Markers are the same as figure~\ref{ecplot} but WKK~4438 is represented by dotted arrows. The lower limits of $kT_\mathrm{e}$ are at 90\% confidence levels. The black solid line represents equation~(\ref{lcomp}). The black dotted line is scaled from the black solid line by a factor of 1/3 in the y-axis direction.}\label{plot}
\end{figure}

\citet{malizia2008} suggested that the accretion rate is a factor determining the electron temperature of a corona. 
We search for any correlation between the accretion rate and the electron temperature of our NLS1 sample. 
We here employ the ratio of the luminosity of seed photons entering the corona ($L_{\mathrm{seed}}$) to the Eddington luminosity ($L_{\mathrm{Edd}}$) to represent the accretion rate. 
Under the assumption that the Compton process is isotropic, a half of the up-scattered photons are emitted to the outside of the corona with a luminosity $L_{\mathrm{Comp}}$, and the other half of them are emitted back to the seed photon field. 
The radiative equilibrium is then expressed as $L_{\mathrm{seed}} = L_{\mathrm{Comp}}$. 
Therefore, the ratio $L_{\mathrm{seed}}/L_{\mathrm{Edd}}$ can be regarded as being equal to $L_{\mathrm{Comp}}/L_{\mathrm{Edd}}$. 
Figure~\ref{plot} shows the plot of $kT_{\mathrm{e}}$ versus $L_{\mathrm{Comp}}/L_{\mathrm{Edd}}$ of our NLS1 sample, where $L_{\mathrm{Comp}}$ was calculated from the flux of the unabsorbed {\tt nthcomp} model. 
Sources with a larger ratio $L_{\mathrm{Comp}}/L_{\mathrm{Edd}}$ tend to have lower $kT_{\mathrm{e}}$ in our sample. 
The NuSTAR results hint that this negative correlation is also seen in the flux variation of SWIFT~J2127.4+5654. 

We try to explain the trend in figure~\ref{plot} by assuming that the dominant channels of electron cooling and heating are inverse Compton scattering and Coulomb collisions with protons, respectively, following a scenario proposed by \citet{miyakawa2008}, who discussed black hole binaries. 
Inverse Compton scattering cools electrons at a rate of 
\begin{equation}
\left(\frac{dE}{dt}\right)_{{\rm Comp}} \sim \frac{4kT_{\mathrm{e}}}{m_{\mathrm{e}}c^2}U_{\mathrm{seed}}n\sigma_{\mathrm{T}}, 
\end{equation}
where $U_{\mathrm{seed}}$ is the seed photon flux density, $n$ is the number density of electrons, and $\sigma_{\mathrm{T}}$ is the Thomson scattering cross section.
The photon flux density is given by $U_{\mathrm{seed}} \simeq L_{\mathrm{seed}}/(\pi r^2)$, where $r$ is the radius of the corona.
Electrons are heated by protons through Coulomb collisions at a rate of
\begin{equation}\label{heat}
\left(\frac{dE}{dt}\right)_{{\rm Coulomb}} \sim \frac{3}{2}\frac{nkT_{\mathrm{p}}}{t_{\mathrm{pe}}}.
\end{equation}
The relaxation time of Coulomb collisions is given by 
\begin{eqnarray}
t_{\mathrm{pe}} &=& \left[\frac{3m_{\mathrm{p}}m_{\mathrm{e}}}{8(2\pi)^{\frac{1}{2}}e^{4}n\ln{\Lambda}}\right] \left(\frac{kT_{\mathrm{p}}}{m_{\mathrm{p}}} + \frac{kT_{\mathrm{e}}}{m_{\mathrm{e}}}\right)^{\frac{3}{2}} \nonumber \\
&\sim& \left[\frac{3m_{\mathrm{p}}m_{\mathrm{e}}}{8(2\pi)^{\frac{1}{2}}e^{4}n\ln{\Lambda}}\right]\left(\frac{kT_{\mathrm{e}}}{m_{\mathrm{e}}}\right)^{\frac{3}{2}},
\end{eqnarray}
which is the time needed for protons to establish a Maxwell distribution \citep{spitzer1962}. 
Here $kT_{\mathrm{p}}$ and $m_{\mathrm{p}}$ denote the proton temperature and the proton rest mass, respectively. 
The Coulomb logarithm $\ln{\Lambda}$ is represented by $\ln{\Lambda} = \ln{\left\{[3/(2e^{3})]\sqrt{(kT_{\mathrm{e}})^{3}/(\pi n)}\right\}}$.
In a corona, the heating rate should balance with the cooling rate, i.e.,
\begin{equation}
\frac{3}{2}\frac{nkT_{\mathrm{p}}}{t_{\mathrm{pe}}} = \frac{4kT_{\mathrm{e}}}{m_{\mathrm{e}}c^2}U_{\mathrm{seed}}n\sigma_{\mathrm{T}}.
\end{equation}
If the proton energy loss rate given by equation~(\ref{heat}) is fast enough compared to the viscous-heating rate through accretion, the proton temperature is approximately given by $kT_{\mathrm{p}} \sim GMm_{\mathrm{p}}/r$. 
As described above, $L_{\mathrm{seed}}$ is comparable with $L_{\mathrm{Comp}}$. 
Thus, $kT_{\mathrm{e}}$ is represented with the ratio $L_{\mathrm{Comp}}/L_{\mathrm{Edd}}$ as
\begin{equation}\label{lcomp}
kT_{\mathrm{e}} = \left[\frac{4m_{\mathrm{p}}\sigma_{\mathrm{T}}}{(2\pi)^{\frac{1}{2}}e^{4}m_{\mathrm{e}}^{\frac{3}{2}}c\tau\ln{\Lambda}}\right]^{-\frac{2}{5}} \left(\frac{L_{\mathrm{Comp}}}{L_{\mathrm{Edd}}}\right)^{-\frac{2}{5}},
\end{equation} 
where $\tau = nr\sigma_{\mathrm{T}}$. 
This equation indicates that $kT_{\mathrm{e}}$ is proportional to $(L_{\mathrm{Comp}}/L_{\mathrm{Edd}})^{-2/5}$ despite a little fluctuation of $\tau$ and $\ln{\Lambda}$. 

We overlay the prediction by equation~(\ref{lcomp}) on our observational results plotted in figure~\ref{plot}. 
In the model calculation, we assumed $\tau=4$ and $r=20r_{\mathrm{g}}$, which gives $n = 2 \times 10^{10}~\mathrm{cm^{-3}}$. 
In spite of the simple approximations, the model curve reproduces the negative slope of $kT_{\mathrm{e}}$ as a function of $L_{\mathrm{Comp}}/L_{\mathrm{Edd}}$ and agrees with the data within a factor of $\sim$ 2--3. 
The factor of $\sim$ 2--3 may indicate that some other cooling mechanisms of electrons have to be taken into account than our very simple assumptions. 

\section{Conclusion}
We studied X-ray spectra of the four NLS1s, Mrk~110, SWIFT~J2127.4+5654, IGR~J16185$-$5928, and WKK~4438 , using Suzaku and NuSTAR data. 
The main conclusions of our work are summarized as follows.
\begin{enumerate}
\item
The spectra of the four sources are reproduced well with a model consisting of a cutoff power law component, a reflection component, and a soft excess component. 
All the four sources gave photon indices of  $\Gamma \sim 2$. 
The cutoff energies were constrained to be $E_{\mathrm{c}}\sim 40~\mathrm{keV}$ for Mrk~110, SWIFT~J2127.4+5654, and WKK~4438 whereas 
a lower limit of $E_{\mathrm{c}} > 155~\mathrm{keV}$ was obtained for IGR~J16185$-$5928. 
Comparing the best-fit values of $\Gamma$ and $E_{\mathrm{c}}$ with those of BLS1s by \citet{malizia2014}, we found that the NLS1s in our sample have systematically softer spectra and lower $E_{\mathrm{c}}$ than the BLS1s. 
Since the electron temperature is related to the cutoff energy as $E_{\mathrm{c}} \approx 2$--$3kT_{\mathrm{e}}$, our result suggests that NLS1s have lower $kT_{\mathrm{e}}$ than BLS1s. 

\item 
We also fitted the Suzaku and NuSTAR spectra, replacing the above mentioned cutoff power law component with a thermal Comptonization model. 
In the case of SWIFT~J2127.4+5654, from which we detected a broad Fe K$\alpha$ emission line, we took into account relativistic reflection on the accretion disk. 
We obtained electron temperatures of $kT_{\mathrm{e}} \sim 10$--$20~\mathrm{keV}$ for Mrk~110 and SWIFT~J2127.4+5654, and lower limits of $kT_{\mathrm{e}} > 18~\mathrm{keV}$ for IGR~J16185$-$5928 and $kT_{\mathrm{e}} > 5~\mathrm{keV}$ for WKK~4438. 
The NuSTAR results hint that $kT_{\mathrm{e}}$ in the high flux period is lower than those in the middle and low flux periods. 
Combining the results of the four sources, we found that $kT_{\mathrm{e}}$ anti-correlates with the ratio $L_{\mathrm{Comp}}/L_{\mathrm{Edd}}$. 
The anti-correlation can be explained by a simple model in which electrons in the corona are cooled and heated  through inverse Compton scattering and Coulomb collisions with protons, respectively.
\end{enumerate}



\begin{ack}
The authors would like to thank to all the members of the Suzaku team and the NuSTAR team. This work is supported by JSPS/MEXT Scientific Research Grant Number JP25109004 (T.T. and T.G.T.), JP15H02090 (T.G.T.), JP26610047 (T.G.T.), JP17K05384 (Y.U.), JP15H02070 (Y.T.), JP16K05296 (Y.T.), and JP23340071 (K.H.).
\end{ack}


\end{document}